\documentclass[12pt,a4paper]{article}

\usepackage[utf8]{inputenc}
\usepackage{amsthm,amsmath,latexsym,amssymb,amsfonts,amscd,cite}
\usepackage{amssymb} 
\usepackage[all]{xy}
\usepackage{tikz}
\usepackage{array}
\usepackage{hyperref}

\usepackage{tocbasic}
\DeclareTOCStyleEntry[
  beforeskip=.2em plus 1pt,
  pagenumberformat=\textbf
]{tocline}{section}

\usepackage{mathrsfs}
\usepackage{hyperref}

\usepackage{epigraph}

\usepackage{bbm}
\usepackage{bm}
\usepackage[T1]{fontenc}
\usepackage{mathrsfs}

\numberwithin{equation}{section}

\theoremstyle{definition}

\newtheorem{example}[equation]{Example}

\usepackage[all]{xy}
\usepackage{graphicx}

\usepackage{mathtools}

\makeatletter
\DeclareRobustCommand\widecheck[1]{{\mathpalette\@widecheck{#1}}}
\def\@widecheck#1#2{%
    \setbox\z@\hbox{\m@th$#1#2$}%
    \setbox\tw@\hbox{\m@th$#1%
       \widehat{%
          \vrule\@width\z@\@height\ht\z@
          \vrule\@height\z@\@width\wd\z@}$}%
    \dp\tw@-\ht\z@
    \@tempdima\ht\z@ \advance\@tempdima2\ht\tw@ \divide\@tempdima\thr@@
    \setbox\tw@\hbox{%
       \raise\@tempdima\hbox{\scalebox{1}[-1]{\lower\@tempdima\box
\tw@}}}%
    {\ooalign{\box\tw@ \cr \box\z@}}}
\makeatother

\newcommand{\sfLambda}{\mathsf{\Lambda}}

\newcommand{\CC}{\mathsf{s}}
\newcommand{\TT}{\mathsf{t}}

\newcommand{\dd}{\mathrm{d}} 

\newcommand{\sfs}{\mathsf{s}} 
\newcommand{\sft}{\mathsf{t}}

\title{{\bf Poisson Electrodynamics With Charged  Matter Fields}}

\date{}
\begin{document}

\maketitle

\begin{center}
\vskip -0.05\textheight
\renewcommand{\thefootnote}{\fnsymbol{footnote}}
  Alexey A. \textsc{Sharapov}\footnote{{\tt sharapov@phys.tsu.ru}} 
\renewcommand{\thefootnote}{\arabic{footnote}}
\setcounter{footnote}{0}

\end{center}
\begin{center}
\emph{Physics Faculty, Tomsk State University,}
\emph{Lenin ave. 36, Tomsk
634050, Russia}
\end{center}

\vspace{5mm}

\begin{abstract}
  Poisson electrodynamics is the low-energy limit of a rank-one noncommutative gauge theory. It admits a closed formulation in terms of a Poisson structure on the space-time manifold and reproduces ordinary classical electrodynamics in the commutative limit. In this paper, we address and solve the problem of  minimal coupling to charged matter fields with a proper commutative limit.  Our construction relies essentially on the geometry of symplectic groupoids and works for all integrable Poisson manifolds. An additional advantage of our approach is that the corresponding Lagrangians can be defined on an arbitrary metric background.
  \end{abstract}

\section{Introduction} 

It is now widely believed that the construction of a viable model of quantum gravity is hardly possible without a radical departure from the concept of a smooth spacetime manifold. The Planck scale ($\ell_P\sim 10^{-35}$ m) is expected to reveal a certain granular structure in the fabric of spacetime. One possible way to model such microscale discreteness is based on ideas of noncommutative geometry. Indeed, if spacetime coordinates $x^\mu$ do not commute, like quantum-mechanical operators, then the generalized uncertainty relation $\Delta x^\mu\,\Delta x^\nu\geq \ell^2_P$  sets an absolute limit to how precisely a spacetime event can be localized. Abandoning spacetime locality can hopefully resolve or at least mitigate the problem of ultraviolet divergences, thereby opening a way to a healthy QFT. All of these ideas go back to the early days of quantum theory \cite{Hei, MBron, Snyder1, Snyder2} and have been discussed sporadically over decades. 

More concrete evidence for spacetime noncommutativity came from string theory. In the seminal paper \cite{Seiberg_1999}, Seiberg and Witten considered the quantization of open strings ending on $D$-branes in the presence of a constant Neveu--Schwartz $B$-field. It was shown that the low-energy effective field theory on a single $D$-brane has a consistent deformation to a noncommutative gauge theory. This result triggered an explosion of research on various aspects of noncommutative field theory, which is reviewed in Refs. \cite{Douglas_2001, Szabo:2001kg, Aschieri_2023, Hersent:2022gry}. In particular, it  became clear that the hopes pinned on noncommutativity were too optimistic: the simplest field-theoretical models on Weyl--Moyal spaces still suffer from divergences due to the phenomenon of UV/IR mixing \cite{IR/UV, UV/IR}. However, one can hope to overcome this problem for more general noncommutative spaces. 

Accepting the hypothesis of spacetime noncommutativity, one might expect the conventional field theories (e.g. the standard model of particle physics) to emerge as effective theories at distances much bigger than the characteristic length scale set by noncommutativity. In other words, the effects of noncommutativity should disappear in the long-wave/low-energy limit, making it legitimate to approximate noncommutative spacetime by a smooth manifold.   The next natural step, suggested by formal analogy with quantum mechanics, is to consider a semi-classical approximation with the fundamental length scale playing the role of a small deformation parameter. In this approximation, the physical spacetime is still modelled by a smooth manifold, while noncommutativity is controlled by an appropriate Poisson structure. The noncommutative algebra generated by spacetime coordinates is expected to arise from the quantization of the underlying Poisson brackets (see the next section).  

In the series of papers \cite{kupriyanov2020non,kupriyanov2021poisson, Kupriyanov_2021,Kupriyanov:2022ohu,Kupriyanov:2023gjj,kupriyanov2023symplectic}, this line of ideas was applied to the gauge vector field and the corresponding gauge theory was dabbed {\it Poisson electrodynamics}. For the best-studied case of Weyl--Moyal spaces, it reproduces the semi-classical limit of noncommutative $U(1)$ gauge theory considered by Seiberg and Witten. Unlike Maxwell's theory, the gauge covariant strength tensor of Poisson electrodynamics is nonlinear and so are the equations  of motion. This places Poisson electrodynamics within a large family of nonlinear modifications of Maxwell's theory, the most notable members of which are   Born--Infeld, Hisenberg--Euler, and ModMax models. Moreover, all of these nonlinear models admit a noncommutative generalization within Poisson electrodynamics. (For a modern account of nonlinear electrodynamics and its applications, we refer the reader to \cite{DSor}.) Some exact solutions to Poisson electrodynamics on the so-called $\kappa$-Minkowski space were found in \cite{Kupriyanov:2022ohu, Kupriyanov:2023gjj}. 

By construction, any Poisson electrodynamics inevitably violates Lorentz invariance because of the presence of a nonzero Poisson tensor. In recent years, the hypothesis of Lorentz invariance violation has gained great interest in both theoretical and experimental high-energy physics, see e.g.~\cite{colladay1998lorentz,coleman1999high,kostelecky2004gravity,kostelecky2009electrodynamics,liberati2013tests,abreu2022testing, Sarker2023InvestigatingTE, Finke_2023}. At low energies, one can  describe Lorentz violating effects by an effective field theory, which was called the  Standard-Model Extension.  This is obtained by extending the standard model Lagrangian with all possible terms that, while violating Lorentz invariance, do respect the standard gauge invariance and renormalization properties.  Various tensor coefficients responsible for breaking Lorentz invariance are interpreted as expectation values in some fundamental (yet unknown) theory. Poisson electrodynamics offers a refined interpretation of these coefficients: all of them arise from a background Poisson tensor. This makes Poisson electrodynamics more specific in its predictions of possible Lorentz violation.

Finally, Poisson electrodynamics features a remarkable duality between the noncommutativity of spacetime and the curvature of momentum space. As shown in \cite{kupriyanov2023symplectic}, the minimal coupling of an electromagnetic field to a point charge implies that the momentum space of the particle is a curved manifold rather than a linear space. The idea of curved momentum space is as old as the idea of non-commutative spacetime \cite{Born, born1949reciprocity} and stems from the same desire to get rid of ultraviolet divergences, see \cite{KOWALSKI_GLIKMAN_2013, Franchino-Vinas:2023rcc} for review. A compact momentum space, for example, would provide an effective UV cut-off for all momentum integrals coming from Feynman diagrams. As we will see in Sec. \S\ref{S2}, the constant Poisson bracket underlying the Weyl--Moyal space does not curve the particle's momentum space, which partly explains the failure of the naive regularization approch by means of constant noncommutativity. At the same time, this indicates the possibility of overcoming the aforementioned UV/IR mixing problem using more general Poisson structures. Even linear Poisson brackets can lead to a compact momentum space. 

This paper can be viewed as a continuation of the recent work \cite{kupriyanov2023symplectic} devoted to the symplectic groupoid approach to Poisson electrodynamics. One of the main results of that work was the general recipe for introducing a minimal coupling of a point charged particle to the electromagnetic field.   In the present paper, we extend this recipe from point particles to charged matter fields in the fundamental representation of $U(1)$.

\section{Noncommutative gauge theory in semi-classical approximation}\label{S1}

Noncommutative field theory proceeds from the fundamental idea that spacetime coordinates are not real numbers, but
noncommuting Hermitian operators $\hat{x}^\mu$ acting in some Hilbert space. In the simplest case, 
one can impose the  following commutation relations:
\begin{equation}\label{CR}
    [\hat{x}^\mu,\hat{x}^\nu]=i a\pi^{\mu\nu}\,.
\end{equation}
 Here $(\pi^{\mu\nu})$ is a constant antisymmetric matrix and $a$ is a parameter of dimension $[{length}]^2$ that measures noncommutativity. Following the analogy with quantum mechanics, one can further assume the commutation relations (\ref{CR}) to arise from the quantization of the constant Poisson brackets
 \begin{equation}\label{PB}
     \{x^\mu,x^\nu\}=\pi^{\mu\nu}\,.
 \end{equation}
In the approach of deformation quantization, for example, this is achieved through the
Weyl--Moyal $\star$-product on the space of functions of the $x$'s.  Explicitly, 
\begin{equation}
    (f\star g) (x)=\exp \left(\frac{ia}{2}\pi^{\mu\nu}\frac{\partial^2}{\partial x^\mu\partial y^\nu}\right)f(x)g(y)|_{y=x}\,.
\end{equation}
Here, the smooth complex-valued functions $f$ and $g$ are viewed as the Weyl symbols of the operators $f(\hat x)$ and $g(\hat x)$.  Let us denote this $\star$-product algebra  by $\mathfrak{A}$. Associated with the algebra $\mathfrak{A}$ is the commutator Lie algebra $\mathrm{L}(\mathfrak{A})$ and the group $\mathrm{G}(\mathfrak{A})$ of the invertible elements of $\mathfrak{A}$.  With some caution, one can view   $\mathrm{L}(\mathfrak{A})$ as the Lie algebra of the infinite-dimensional Lie group  $\mathrm{G}(\mathfrak{A})$. 
Regarding the elements of $\mathfrak A$ as scalar fields and replacing the pointwise product of functions with the $\star$-product above, it is not hard to formulate noncommutative counterparts
of various field-theoretical models \cite{Douglas_2001, Szabo:2001kg, Aschieri_2023,Hersent:2022gry}.

In this paper, we are interested in the noncommutative generalization of classical electrodynamics with complex matter fields. 
As usual, the electromagnetic field is described by the gauge potential $A_\mu(x)$, whose components are now viewed as elements of $\mathfrak{A}$, and the gauge group is postulated to be $\mathrm{G}(\mathfrak{A})$. The expressions for the gauge covariant strength tensor
\begin{equation}\label{Fst}
    F_{\mu\nu}=\partial_\mu A_\nu-\partial_\nu A_\mu+\frac{g}{ia}[A_\mu,A_\nu]_{\star}
\end{equation}
and the gauge transformations
\begin{equation}\label{GT}
    A_\mu\mapsto A_\mu^U=U\star A_\mu\star U^{-1}+\frac{ia}{g}U\star \partial_\mu U^{-1}\quad \Rightarrow\quad F_{\mu\nu}\mapsto F_{\mu\nu}^U=U\star F_{\mu\nu}\star U^{-1}
\end{equation}
just mimic  those from  Yang--Mills theory. Here $g$ is a coupling constant of dimension $[ length ]$, $U$ is an arbitrary element of $\mathrm{G}(\mathfrak{A})$, and $[f,h]_\star:=f\star h-h\star f$. Setting $U=1+(g/ia)\varepsilon$ for an infinitesimally small $\varepsilon \in \mathfrak{A}$, we can rewrite (\ref{GT}) as
\begin{equation}\label{gtr}
    \delta_\varepsilon A_\mu =\partial_\mu\varepsilon +\frac{g}{ia}[A_\mu,\varepsilon]_\star \quad \Rightarrow\quad \delta_\varepsilon F_{\mu\nu}=
    \frac{g}{ia}[F_{\mu\nu}, \varepsilon ]_\star\,.
    \end{equation}
The commutator of two infinitesimal gauge transformations reads
\begin{equation}\label{galg}
[\delta_{\varepsilon},\delta_{\varepsilon'}]=\delta_{\frac{g}{ia}[\varepsilon,\varepsilon']_\star}\,.
\end{equation}
This means that the Lie algebra of gauge transformations is isomorphic to $\mathrm{L}(\mathfrak{A})$.

The strength tensor (\ref{Fst}) is a building block for constructing a gauge-invariant action functional and field equations.  Let $\Phi(x)$ denote a complex field that may be multi-component. Similar to Yang--Mills  theory,  each component of $\Phi$ is assumed to transform under a linear representation of the gauge group $\mathrm{G}(\mathfrak{A})$. 
The origin of the group $\mathrm{G}(\mathfrak{A})$ suggests three natural representations in the space $\mathfrak A$, namely, left regular, right regular, and adjoint.
These lead to the following gauge transformations of the field $\Phi$: 

$$
\Phi\mapsto U\star \Phi \,, \qquad \Phi\mapsto \Phi\star U^{-1} \,,\qquad \Phi\mapsto U\star \Phi \star U^{-1}\,.
$$

The minimal interaction between the electromagnetic and charged matter fields is introduced through the gauge covariant derivative. 
The form of the covariant derivative depends on a representation carried by $\Phi$. For the three representations above we have, respectively, 
$$
D^{\mathrm l}_\mu\Phi=\partial_\mu\Phi +\frac{g}{ia}A_\mu\star \Phi\,,\qquad D^{\mathrm r}_\mu\Phi=\partial_\mu\Phi -\frac{g}{ia} \Phi\star A_\mu\,,\qquad   D^{\mathrm{ad}}_\mu\Phi=\partial_\mu\Phi +\frac{g}{ia}[A_\mu, \Phi]_\star\,.
$$
There are two interesting limits to consider: the limit of commutativity, $\pi^{\mu\nu}=0$, and the semi-classical limit  $a\rightarrow 0$. 
In the former case,  the strength tensor (\ref{Fst}) reduces to the conventional electromagnetic field tensor, the left and right covariant derivatives go into the standard covariant derivative of $U(1)$ gauge theory, and we arrive at ordinary electrodynamics with electric charge $e=\pm g/a$.  For the adjoint representation, the commutative limit  amounts  to the free limit $g\rightarrow 0$. 

As to the semi-classical limit $a\rightarrow 0$, it makes sense only for  matter fields in the adjoint representation. 
In this limit, the $\star$-commutators above turn into the Poisson brackets defined by Eq. (\ref{PB}) and the expressions for the strength tensor, gauge transformations, and covariant derivative take the following form \cite{kupriyanov2021poisson}:
\begin{equation}\label{PED}
\begin{array}{ll}
   F_{\mu\nu}=\partial_\mu A_\nu-\partial_\nu A_\mu+{g}\{A_\mu,A_\nu\}\,,&\qquad  \delta_\varepsilon A_\mu=\partial_\mu\varepsilon +{g}\{A_\mu,\varepsilon\}\,,\\[3mm] 
   D^{\mathrm{ad}}_\mu\Phi=\partial_\mu\Phi +{g}\{A_\mu, \Phi\}\,,&\qquad \delta_\varepsilon \Phi =-g\{\varepsilon, \Phi\}\,.
   \end{array}
   \end{equation}
From the physical viewpoint, the semi-classical approximation is nothing else but the long-wave limit for the electromagnetic field. Indeed, if $A_\mu(x)=A_{0\mu}e^{\frac{2\pi i x}{\lambda}}$ is an electromagnetic wave with $\lambda^2>>a$, one can safely omit all higher-derivative terms in the strength tensor (\ref{Fst}), keeping only the first-order derivatives. This gives immediately the semi-classical strength tensor (\ref{PED}).  

It follows from Eq. (\ref{PED}) that the semi-classical limit of the gauge algebra $\mathrm{L}(G)$ is given by the Lie algebra of functions with respect to the Poisson bracket, that is, 
\begin{equation}\label{PED1}
    [\delta_{\varepsilon_1},\delta_{\varepsilon_2}]=\delta_{g\{\varepsilon_1, \varepsilon_2\}}\,.
    \end{equation}
Let us denote this algebra by $\mathcal{L}$. The Lie algebra $\mathcal{L}$ is still noncommutative unless  $g\neq 0$. It is remarkable that the long-wave limit of the noncommutative electrodynamics with  matter in adjoint is described by exact (not approximate) and self-consistent formulas (\ref{PED}, \ref{PED1}). 
In \cite{kupriyanov2023symplectic}, a gauge theory with the underlying gauge algebra (\ref{PED1}) determined by a (not necessary constant) Poisson bracket was dabbed {\it Poisson electrodynamics} (PED).  Redefining $\pi^{\mu\nu}\rightarrow g\pi^{\mu\nu}$ we can remove the coupling  constant $g$ from (\ref{PED}, 
\ref{PED1}). After this redefinition, the weak coupling limit of Poisson electrodynamics is recovered as the commutative limit.

If we now accept the hypothesis that spacetime is truly noncommutative, i.e., $\pi^{\mu\nu}\neq 0$ and $a>0$, then
the ordinary Maxwell's electrodynamics have to arise as the long-wave ($\lambda^2>> a$) and weak coupling ($g\rightarrow 0$) regime of its noncommutative counterpart or, what is the same, as the commutative limit of PED.  Such an interpretation, however, runs immediately into the following problem. For the weak coupling regime,  we would expect all formulas of classical electrodynamics to be restored as $g\rightarrow 0$, including the formula for the $U(1)$ gauge transformations of charged matter fields. 
We refer to such a behaviour as the {\it correspondence principle}.  On the other hand, Eq. (\ref{PED}) for the gauge transformations of $\Phi$ has nothing to do with the standard $U(1)$ gauge transformation and the matter field completely decouples from the electromagnetic field as $g\rightarrow 0$. This goes against the correspondence principle. With the $U(1)$ gauge group in mind, one could try the standard gauge transformations
\begin{equation}\label{abgt}
\delta_\varepsilon \Phi= -i\varepsilon\cdot \Phi, 
\end{equation}
replacing the adjoint action. Such a replacement, however, does not work for a very simple reason. Whenever $g\neq 0$ the infinitesimal gauge transformations, when evaluated on $A_\mu$, form a nonabelian 
Lie algebra (\ref{PED1}), while the gauge transformations (\ref{abgt}) commute.  Therefore we should consider more elaborate gauge algebra generators  for the matter fields. It seems  the only way  to reconcile the  limiting abelian transformation  (\ref{abgt}) with the  nonabelian gauge algebra  (\ref{PED1}) is to admit the dependence of the gauge field $A$.\footnote{It might be worth  noting that geometric quantization offers the following representation of the Lie algebra (\ref{PED1}): $\delta_\varepsilon \Phi =(\varepsilon  +x^\mu\partial\varepsilon/\partial x^\mu)\Phi +g\{\varepsilon,\Phi\}$. This is known as the Kostant--Souriau prequantization \cite[Ch. 8]{woodhouse1992geometric}. As for the adjoint action, this representation does not have a proper commutative limit.} So, we replace (\ref{abgt}) with a more general expression
\begin{equation}\label{gg}
    \delta_{\varepsilon}\Phi=-i\alpha(A, \varepsilon)\cdot \Phi\,,
\end{equation}
where the generator of gauge transformations $\alpha (\varepsilon, A)$ 
satisfies the `boundary condition'
\begin{equation}\label{bc}
    \alpha(A, \varepsilon)=\varepsilon +\mathcal{O}(g)\,.
\end{equation}
The commutation relations (\ref{PED1}) lead then to the equation
\begin{equation}\label{1c}
    \delta_{\varepsilon_1}\alpha(A, \varepsilon_2)-  \delta_{\varepsilon_2}\alpha(A, \varepsilon_1)= \alpha(A, g\{\varepsilon_1,\varepsilon_2\})\,.
\end{equation}
This equation has a nice mathematical interpretation. 
Let $\mathcal{A}$ denote the space of functionals of  the  fields $A_\mu$ with values in smooth functions of the  $x$'s.  The gauge transformations (\ref{PED}) give $\mathcal{A}$ the structure of a module over the Lie algebra $\mathcal{L}$. Then Eq. (\ref{1c}) defines a $1$-cocycle $\alpha$ of the Lie algebra $\mathcal{L}$ with coefficients in $\mathcal{A}$. The cocycle condition always admits trivial solutions given by $1$-coboundaries 
\begin{equation}
    \alpha(A,\varepsilon)=\delta_\varepsilon \beta(A)\qquad \forall \beta \in \mathcal{A}\,.
\end{equation}
All these solutions, however, violate the boundary condition (\ref{bc}). 
Notice that $\delta_\varepsilon A_i=0$ whenever $\varepsilon$ is equal to some constant $c$.  
Then it follows from (\ref{bc}) that 
\begin{equation}
    0= \delta_c \beta(A)=\alpha(A, c)\neq c+\mathcal{O}(g)\,,
\end{equation}
a contradiction. 
Therefore, we are interested in solutions to (\ref{1c}) that are given by nontrivial $1$-cocycles $\alpha$. The existence of such solutions is not obvious in advance. However, one can verify that the expression
\begin{equation}\label{ans}
\begin{array}{c}
    \alpha(A, \varepsilon)=\varepsilon +\frac 12 g\pi^{\mu\nu}A_\mu\partial_\nu\varepsilon+\mathcal{O}(g^2) 
    \end{array}
\end{equation}
satisfies (\ref{1c}) up to the first order in $g$. Let us assume for a moment that the ansatz (\ref{ans}) extends to all orders in $g$. 
The next question to ask is about a covariant derivative that respects the gauge transformations (\ref{gg}).  
Again, we can try the following natural ansatz:
\begin{equation}
    D_\mu\Phi=\partial_\mu\Phi +\theta_\mu(A)\Phi\,.
\end{equation}
The correspondence principle implies that $\theta_\mu(A)= A_\mu+\mathcal{O}(g)$. Verifying the covariance condition $\delta_\varepsilon D_\mu\Phi=D_\mu\delta_\varepsilon\Phi$, we obtain
\begin{equation}\label{tt}
\delta_\varepsilon \theta_\mu(A)=\partial_\mu\alpha(A, \varepsilon)\,.
\end{equation}
Up to the first order in $g$ one can satisfy this equation by setting 
\begin{equation}\label{tA}
\begin{array}{c}
    \theta_\mu(A)=A_\mu +g \pi^{\nu \gamma} A_\nu\big( \partial_\gamma A_\mu -\frac12 \partial_\mu A_\gamma\big) +\mathcal{O}(g^2)\,.
    \end{array}
\end{equation}
Thus, unlike ordinary electrodynamics, the connection $1$-form $\theta=\theta_\mu(A)\dd x^\mu$ is not just the gauge potential $A=A_\mu(x)\dd x^\mu$; rather, it is a composite field of $A$.

A glance is enough to recognize in formulas (\ref{ans}) and (\ref{tA}) the leading terms of the Seiberg--Witten map for a rank-one gauge theory \cite{Seiberg_1999}, \cite{Aschieri_2023}, \cite{2023JPhA...56K5201K}. 
By the very logic, these formulas should arise as the semi-classical limit of the Seiberg--Witten map relating the abelian gauge transformations $\delta_{\varepsilon}A=\dd \varepsilon$ with non-abelian ones (\ref{gtr}).  Such a limit has been discussed, for example, in \cite{jurvco2001nonabelian}. The existence of the Seiberg--Witten map ensures the existence of an all-order solution to Eqs. (\ref{1c}) and (\ref{tt}).

In the next sections, we will generalize the above considerations in two directions. First, we will replace the constant Poisson brackets (\ref{PB}) by an arbitrary Poisson structure. Second, we will explain the geometric origin of the above formulas and define  the group of {\it finite} gauge transformations, whose infinitesimal version is given by Eqs. (\ref{PED}, \ref{gg}). 
 
For the case of a pure electromagnetic field, the first generalization was proposed some time ago in   \cite{kupriyanov2020non}, see also \cite{Kupriyanov_2021}. 
The gauge transformations were postulated  in the form
\begin{equation}\label{igtr1}
\delta_\varepsilon A_\mu=\gamma_\mu^\nu (x,A)\,\partial_\nu\varepsilon +\{A_\mu, \varepsilon\}\,,
\end{equation}
where the braces  now denote an arbitrary Poisson bracket on the spacetime manifold and the functions $\gamma_\mu^\nu(x,A)$ are determined from the commutation relations  (\ref{PED1}). The geometric interpretation of formula (\ref{igtr1}) in terms of {\it symplectic groupoids} was proposed in the recent paper \cite{kupriyanov2023symplectic}. According to this interpretation, the gauge fields $A$ are identified with bisections of a symplectic groupoid integrating a given Poisson structure; hence, they assume values in some nontrivial manifold rather than a vector space. The linear superposition principle for the free electromagnetic field is replaced with a nonlinear one. The latter owes its existence to the fact that the bisections form a non-abelian group and can be multiplied in an associative way. In terms of this group structure, the gauge transformations (\ref{igtr1}) come from left translations by the elements of the subgroup of Lagrangian bisections. Furthermore, this geometric interpretation offers a simple way to introduce a minimal coupling of a point charge to external electromagnetic fields if one identifies the phase space of the charged particle with the total space of the symplectic groupoid. In the next sections, we will extend the construction of minimal interaction from classical particles to complex matter fields. 

\section{Electromagnetic fields on Poisson manifolds}\label{S2}

We start from the assumption that the semi-classical limit of a noncommutative spacetime $X$ is described by some Poisson brackets of local coordinates,
\begin{equation}\label{PB1}
    \{x^\mu,x^\nu\}=\pi^{\mu\nu}(x)\,.
\end{equation}
The Jacobi identity implies the following equations for the bivector $\pi^{\mu\nu}$:
\begin{equation}\label{JI}
    \pi^{\mu \,\lambda}\partial_\lambda\pi^{\nu\gamma }+ \pi^{\nu\lambda}\partial_\lambda\pi^{\gamma\mu}+  \pi^{\gamma\lambda}\partial_\lambda\pi^{\mu \nu}=0\,.
\end{equation}
Regarding now the variables $x^\mu$ as the position coordinates of a point particle, one can wonder about the corresponding phase space, that is, the space of states of the particle. Let us denote this hypothetical phase space by $\mathcal{G}$.  With the correspondence principle in mind, we can assume
that the dimension of the phase space is twice the dimension of the configuration space, as in ordinary mechanics. 
This, however, says  nothing about the symplectic structure or topology of $\mathcal G$. Therefore, we may consider the most general Poisson brackets completing (\ref{PB1}) to a symplectic structure:
\begin{equation}\label{PB2}
    \{x^\mu, p_\nu\}=\gamma^\mu_\nu(x,p)\,,\qquad \{p_\mu, p_\nu\}=\Delta_{\mu\nu}(x,p)\,.
\end{equation}
Here the new variables $p_\mu$ are the `momenta conjugated to the position coordinates $x^\mu$'. The functions $\gamma$ and $\Delta$ are to be determined from the Jacobi identity. It is also assumed that the Poisson brackets (\ref{PB1}, \ref{PB2}) are nondegenerate to define a symplectic structure on $\mathcal G$.  Evaluating the Jacobi identities, one obtains a set of nonlinear first-order PDEs for the structure functions $\gamma$ and $\Delta$. 
These equations are known to be solvable at least locally.  For example, one can always set $\Delta_{\mu\nu}=0$ and look for $\gamma^\mu_\nu$ in the form of power series in momenta:
\begin{equation}
\gamma^\mu_\nu(x,p)=\delta^\mu_\nu+\sum_{k=1}^\infty \gamma^{\mu\mu_1\cdots \mu_k}_\nu(x)p_{\mu_1}\cdots p_{\mu_k}\,.
\end{equation}
Upon substituting this expansion in the Jacobi identity
\begin{equation}
    \{x^\mu, \{x^\nu, p_\lambda\}\}+ \{x^\nu, \{p_\lambda, x^\mu\}\}+ \{p_\lambda, \{x^\mu, x^\nu\}\}=0\,,
    \end{equation}
one can recurrently express the functions $\gamma^{\mu\mu_1\cdots \mu_k}_\nu(x)$ through the Poisson bivector $\pi^{\mu\nu}(x)$ and its derivatives \cite{kupriyanov2019recurrence, Kupriyanov_2021}.  Moreover, the series converges  in a tubular neighbourhood of $X\subset T^\ast X$, with $p_i$ being coordinates in the cotangent spaces. It is these functions $\gamma^\mu_\nu(x,p)$ that determine the infinitesimal gauge transformations (\ref{igtr1}).  This indicates a deep connection between the noncommutative kinematics of  a point particle and Poisson electrodynamics.

Of course, it would be naive to identify the phase space in question with the total space of the cotangent bundle $T^\ast X$; at best, the latter describes only some part of $\mathcal G$. To reconstruct the whole phase space as a smooth manifold, more data is needed. In the recent paper \cite{kupriyanov2023symplectic}, 
this missing part of the data was identified with the structure of a {\it symplectic groupoid}
integrating a given Poisson manifold $(X,\pi)$.  In a nutshell, this means that, like the cotangent bundle $T^\ast X$, the true phase space $\mathcal{G}$ is a fiber bundle over $X$ with two important differences: (i) a fibre  -- the space of momenta -- is now allowed to be a nontrivial manifold (rather than a linear space) and (ii) there are {\it two} canonical projections from $\mathcal G$ to $X$.\footnote{The existence of two canonical projections is stressed in the shorthand notation $\mathcal{G}\rightrightarrows X$ for a groupoid $\mathcal G$ over  $X$.}  Both of these distinctions are due to the noncommutativity of spacetime and disappear in the commutative limit. The relation between the configuration space $X$ and a (nonlinear) momentum space is roughly the same as between Lie algebras and Lie groups\footnote{The existence of a Lie group integrating a given Lie algebra is the content of the Cartan--Lie theorem. In contrast, not every Poisson manifold comes from (or can be integrated to) a smooth symplectic groupoid. Fortunately, all integrability conditions have now been established \cite[Ch.14]{CFM} and are assumed to be satisfied below. Such Poisson manifolds are often called {\it integrable}.

}. Example \ref{Exam3} below shows that this is more than just an analogy.

Having settled the issues  of noncommutative kinematics, we are ready to formulate the basic principles of Poisson electrodynamics. There are four of them \cite{kupriyanov2023symplectic}:

\begin{itemize}
    \item[P1)] The physical spacetime is a smooth manifold $X$ endowed with a Poisson bivector $\pi$. 
    \item[P2)] The phase space of a point charged particle on $X$ is a symplectic groupoid $\mathcal{G}\rightrightarrows X$ integrating the Poisson manifold $(X,\pi)$.
    \item[P3)] The configuration space of the electromagnetic field is identified with the group  of bisections $\mathscr{B}(\mathcal{G})$ of the symplectic groupoid $\mathcal{G}\rightrightarrows X$. 
    \item[P4)] The gauge group of the electromagnetic field is given by the subgroup of Lagrangian bisections $\mathscr{L}(\mathcal{G})\subset 
    \mathscr{B}(\mathcal{G})$; the group $\mathscr{L}(\mathcal{G})$ acts on $\mathscr{B}(\mathcal{G})$ by right translations. 
    \end{itemize}

\paragraph{Remark.} The brief comments below are primarily intended to set our notation rather than to define or explain symplectic groupoids in any detail.   A comprehensive discussion of the geomtry of symplectic groupoids can be found in\cite{KM, WdS, MC, CFM}. For a quick introduction to the subject, we refer the reader to  \cite[\S 2]{kupriyanov2023symplectic} 

As the name suggests, the total space of a symplectic groupoid $\mathcal{G}$ is  a symplectic manifold. By definition, $\dim \mathcal{G}=2\dim X$. Let  $\omega$ denote the 
corresponding symplectic $2$-form. The bivector dual to $\omega$  endows $\mathcal{G}$ with a Poisson structure. 
The  pair of aforementioned projections $\sfs, \sft: \mathcal{G}\rightarrow X$, entering the definition of a symplectic groupoid, 
are called {\it source} and {\it target}. Either of them makes $\mathcal{G}$ into a fibre bundle over $X$. It is required that  $\CC$ be Poisson and $\TT$ be anti-Poisson map:
\begin{equation}\label{sstt}
   \{\CC^\ast f, \CC^\ast g\}_{\mathcal{G}}= \{f,g\}_X\,,\qquad   \{\TT^\ast f, \TT^\ast g\}_{\mathcal{G}} =-\{f,g\}_X \qquad \forall f,g\in C^\infty(X)\,.     \end{equation}
This allows one to refer to the source map  $\sfs: \mathcal{G}\rightarrow X$ as a {\it symplectic realization} of the Poisson manifold $(X,\pi)$. The sets $\sfs^{-1}(x)$ and $\sft^{-1}(x)$ are called the source and target fibers over $x\in X$. The definition of a symplectic groupoid also includes a smooth involution  $\mathsf{i}: \mathcal{G}\rightarrow \mathcal{G}$, called inversion, that has $X$ as a fixed point set and maps $s$-fibres to $t$-fibers. More precisely,  $\mathsf{t}=\mathsf{s}\circ\mathsf{i}$ and $\mathsf{i}\circ\mathsf{i}=\mathrm{id}_{\mathcal{G}}$.

Associated with every fibre bundle is the set  of smooth sections. By definition, a submanifold $\Sigma\subset \mathcal{G}$ is called a {\it bisection} if both restrictions $\CC|_\Sigma$ and $\TT|_\Sigma$ are diffeomorphisms on the base manifold $X$. It is clear that $\dim \Sigma=\dim X$. Each bisection $\Sigma$ defines the pair of maps 
\begin{equation}
    \Sigma_{\sfs}: X\rightarrow \mathcal{G}\,,\qquad \Sigma_{\sft}: X\rightarrow \mathcal{G}
\end{equation}
that are specified by the following properties: 
\begin{equation}
    \mathrm{Im}\,\Sigma_{\sfs}=\Sigma=\mathrm{Im}\,\Sigma_{\sft}\quad\text{and}\quad \sfs \circ \Sigma_\sfs=\mathrm{id}_X= \sft \circ \Sigma_\sft \,.  
\end{equation}

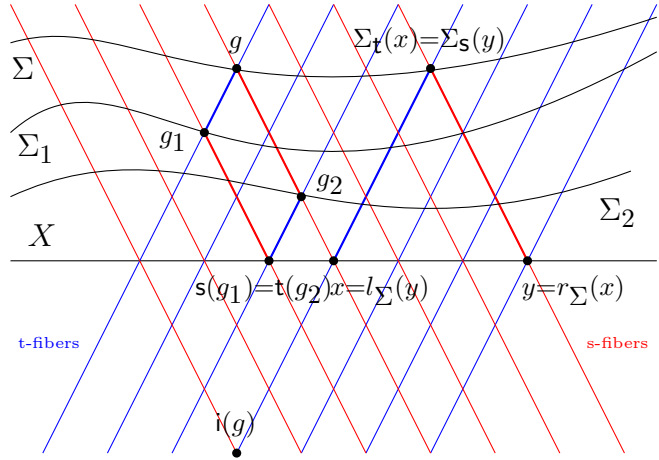
\begin{figure}[h]
\center{
\Large
\begin{tikzpicture}[scale=1.7]

\draw[-, thick, color=blue] (9,0) -- (9.75,1.5);
\draw[-, thick, color=red] (9.75,1.5) -- (10.5, 0);

\draw[-, thick, color=blue] (8,1) -- (8.25,1.5);
\draw[-, thick, color=blue] (8.5,0) -- (8.75, 0.5);

\draw[-, thick, color=red] (8.5,0) -- (8,1);
\draw[-, thick, color=red] (8.75,0.5) -- (8.25, 1.5);

\draw[] (6.5,0) -- (11.5,0);
\draw[-, color=blue] (9.75,-1.5) -- (11.5,2);
\draw[-, color=blue] (7.25,-1.5) -- (9,2);
\draw[-, color=blue] (8.25,-1.5) -- (10,2);
\draw[-, color=blue] (9.25,-1.5) -- (11,2);
\draw[-, color=blue] (6.75,-1.5) -- (8.5,2);
\draw[-, color=blue] (7.75,-1.5) -- (9.5,2);
\draw[-, color=blue] (8.75,-1.5) -- (10.5,2);

\draw[-, color=red] (9.5,2) -- (11.25,-1.5);
\draw[-, color=red] (7,2) -- (8.75,-1.5);
\draw[-, color=red] (8,2) -- (9.75,-1.5);
\draw[-, color=red] (9,2) -- (10.75,-1.5);
\draw[-, color=red] (6.5,2) -- (8.25,-1.5);
\draw[-, color=red] (7.5,2) -- (9.25,-1.5);
\draw[-, color=red] (8.5,2) -- (10.25,-1.5);

\fill [black]  (8.5,0) circle (1pt);
\fill [black]  (8,1) circle (1pt);
\fill [black]  (8.75,0.5) circle (1pt);
\fill [black]  (8.25, 1.5) circle (1pt);
\fill [black]  (9.75,1.5) circle (1pt);
\fill [black]  (9,0) circle (1pt);
\fill [black]  (10.5,0) circle (1pt);

\fill [black]  (8.25,-1.5) circle (1pt);

\coordinate [label=below:$\!{}_{{}_{\sfs(g_1)=\sft(g_2)}}$] (A) at (8.5, 0);

\coordinate [label=above:${}_{{}_{\Sigma_{\sft}(x)=\Sigma_{\sfs}(y)}}$] (A) at (9.75, 1.5);
\coordinate [label=right:${}_{{}_{g_2}}$] (A) at (8.75, 0.6);
\coordinate [label=left:${}_{{}_{g_1}}$] (A) at (8, 0.93);
\coordinate [label=above:${}_{{}_{g}}$] (A) at (8.25, 1.5);
\coordinate [label=below:${}_{{}_{\qquad\; x=l_{\Sigma}(y)}}$] (A) at (9, 0);
\coordinate [label=below:${}_{{}_{\qquad\; y= r_{\Sigma}(x)}}$] (A) at (10.5, 0);

\coordinate [label=above:${}_{{}_{\mathsf{i}(g)}}$] (A) at (8.25, -1.5);

\coordinate [label=:${}_{X}$] (A) at (6.75, 0);
\coordinate [label=below:${}_{\Sigma_1}$] (A) at (6.7, 1.1);
\coordinate [label=below:${}_{\Sigma_2}$] (A) at (11.2, 0.6);

\coordinate [label=below:${}_{\Sigma}$] (A) at (6.6, 1.7);

\coordinate [label=:${}_{\textcolor{blue}{\text{\tiny t-fibers}}}$] (A) at (6.8, -0.8);
\coordinate [label=:${}_{\textcolor{red}{\text{\tiny s-fibers}}}$] (A) at (11.2, -0.8);

\draw [] (6.5, 1.7) ..controls (7.5, 2)  and (8.2, 0.82).. (11.5,1.9);
\draw [] (6.5, 1) ..controls (7.5, 1.9)  and (8.2,-0.2).. (11.5,1.63);
\draw [] (6.5, 0.5) ..controls (8, 1.2)  and (9,-0.15).. (11.3,0.7);
\end{tikzpicture}
}
\caption{\small The multiplication of bisections and their action on the base manifold. }
\label{fig}
\end{figure}

Let $\mathscr {B}(\mathcal{G})$ denote the set of all bisections of the groupoid $\mathcal{G}\rightrightarrows X$. Significantly, $\mathscr{B}(\mathcal{G})$ comes equipped with a natural group structure. The group product is defined by the `parallelogram rule': If $\Sigma_1, \Sigma_2$ is a pair of bisections, then for any points $g_1\in \Sigma_1$ and $g_2\in \Sigma_2$ such that $\CC(g_1)=\TT(g_2)$ there exists a unique point $g$ satisfying
$\sft(g)=\sft(g_1)$ and $\sfs(g)=\sfs(g_2)$. By definition, the point $g$ belongs to the product 
$\Sigma=\Sigma_1\Sigma_2$,
see Fig. \ref{fig}. The role of the identity element of the group $\mathscr {B}(\mathcal{G})$ is played by the base manifold $X$, that is, $X\Sigma=\Sigma=\Sigma X$ and the inversion is defined as $\Sigma^{-1}=\mathsf{i}(\Sigma)$.

The group $\mathscr B(\mathcal{G})$ acts naturally on the base manifold $X$ by diffeomorphisms:
\begin{equation}\label{rl}
    x\mapsto l_\Sigma(x)=\sft \circ \Sigma_\sfs(x)\,,\qquad x\mapsto r_\Sigma(x)=\sfs \circ \Sigma_\sft(x)\qquad \forall x\in X \quad \forall\Sigma\in \mathscr{B}(\mathcal{G})\,,
    \end{equation}
see Fig. (\ref{fig}). It follows from the definition that $r_\Sigma=l_\Sigma^{-1}$ and
\begin{equation}
    l_{\Sigma_1\Sigma_2}=l_{\Sigma_1}l_{\Sigma_2}\,,\qquad 
    r_{\Sigma_1\Sigma_2}=r_{\Sigma_2}r_{\Sigma_1}    \,.
    \end{equation}
Notice that $l_\Sigma=\mathrm{id}_X=r_\Sigma$ whenever $\sfs=\sft$. 

Compatibility between the groupoid and symplectic structures on $\mathcal{G}$ is expressed by the following {\it multiplicative properties} of $\omega$:
\begin{equation}\label{mult}
(\Sigma_1\Sigma_2)^\ast_\sfs \omega= (\Sigma_2)^\ast_\sfs \omega+l^\ast_{\Sigma_2}(\Sigma_1)_\sfs^\ast\omega\,,\qquad (\Sigma_1\Sigma_2)^\ast_\sft \omega= (\Sigma_1)^\ast_\sft \omega+r^\ast_{\Sigma_1}(\Sigma_2)_\sft^\ast\omega\,.
\end{equation}

A bisection $\Sigma$ is called {\it Lagrangian} if $\omega|_\Sigma=0$. It follows from (\ref{mult}) that the  Lagrangian bisections form a subgroup $\mathscr{L}(\mathcal{G})$ in the group $\mathscr{B}(\mathcal{G})$. In particular, $\omega|_{X}=0$.

\begin{example}\label{Exam1}({\it Zero Poisson structures}.) Conventional electrodynamics in commutative spacetime $X$ corresponds to the case $\pi=0$. The cotangent bundle of $X$ endowed with the canonical symplectic form gives a symplectic groupoid integrating the zero Poisson structure. In other words,  $\mathcal{G}=T^\ast X$ and $\omega = \dd \vartheta$, where  $\vartheta=p_\mu\dd x^\mu$ is the Liouville  $1$-form on $T^\ast X$. The source and target maps coincide with each other and with the canonical projection $\mathsf{p}: T^\ast X\rightarrow X$. As a result, the bisections are just sections of the cotangent bundle  of $X$ and the group $\mathscr {B}(\mathcal{G})$ is given by the additive group $\sfLambda^1(X)$ of $1$-forms on $X$. Since $\sfs=\mathrm{p}=\sft$, the group $\mathscr {B}(\mathcal{G})$ acts trivially on $X$. The (bi)section  $p_\mu=A_\mu(x)$ defined by a $1$-form $A=A_\mu(x)\dd x^\mu$ is Lagrangian iff the form $A$ is closed, $\dd A=0$.  Hence, the group $\mathscr{L}(\mathcal{G})$ consists of closed $1$-forms. These form a subgroup $\mathsf{Z}\sfLambda^1(X)\subset \sfLambda^1(X)$. If ${H}^1(X)=0$, e.g. the manifold $X$ is simply connected, then every closed $1$-form is exact and $\mathsf{Z}\sfLambda^1(X)=\dd \sfLambda^0(X)$.  
Identifying now the $1$-forms $A$ with electromagnetic potentials of Maxwell's electrodynamics, we arrive at the standard gauge transformations:
\begin{equation}
    A\mapsto A'=A+\dd f\,,\qquad \forall f\in \sfLambda^0(X)\,.
\end{equation}

\end{example}

\begin{example}\label{Exam2} ({\it Constant Poisson structures.}) The next to the trivial Poisson structure is the class of constant Poisson structures. These are defined in Cartesian space $X=\mathbb{R}^n$ with coordinates $x^\mu$. The corresponding Poisson brackets 
\begin{equation}\label{Exx}
    \{x^\mu,x^\nu\}=\pi^{\mu\nu}
\end{equation}
are determined by a constant antisymmetric matrix $\pi^{\mu\nu}$. This Poisson structure extends to the space $\mathbb{R}^{2n}=\mathbb{R}^n\times \mathbb{R}^n$ by adding new coordinates $p_\mu$ together with the Poisson brackets 
\begin{equation}\label{Epp}
    \{ x^\mu, p_\nu\}=\delta^\mu_\nu\,,\qquad \{p_\mu,p_\nu\}=0\,.
\end{equation}
Together Eqs. (\ref{Exx}) and (\ref{Epp}) define a nondegenerate  Poisson structure on $\mathbb{R}^{2n}$. The corresponding symplectic form is given by 
\begin{equation}\label{E2SS}
\begin{array}{c}
    \omega=\dd p_\mu\wedge \dd x^\mu+\frac12 \pi^{\mu\nu}\dd p_\mu\wedge \dd p_\nu\,.
    \end{array}
\end{equation}
A compatible groupoid structure is determined by the projections 
\begin{equation}\label{E2ST}
    \sfs(p_\nu,x^\mu)=x^\mu\,, \qquad \sft (p_\nu, x^\mu)=x^\mu+\pi^{\mu\nu}p_\nu\,.
\end{equation}
One can easily verify relations (\ref{sstt}). Bisections $\Sigma=(A_\nu(x), x^\mu)\subset \mathbb{R}^n\times\mathbb{R}^n$ are defined as graphs of smooth maps
$A: \mathbb{R}^n\rightarrow \mathbb{R}^n$. The next formula gives the composition of two bisections:
\begin{equation}\label{AB}
\big(A_\mu(x), x^\nu\big)\big(B_\mu(x),x^\nu\big)=\big(A_\mu\big(x^\lambda+\pi^{\lambda\gamma}B_\gamma(x)\big)+B_\mu(x),\, x^\nu\big)\,.
\end{equation}
By definition, $X=(0,x)\subset \mathbb{R}^n\times \mathbb{R}^n$ and the inversion formula reads
\begin{equation}
       \big(A_\mu(x), x^\nu\big)^{-1}=\big(-A_\mu(x),\, x^\nu+\pi^{\nu\lambda}A_\lambda(x)\big)\,.
       \end{equation}
A Lagrangian bisection $\Sigma=(B_\mu(x),x^\nu)$ is defined by the equation 
\begin{equation}
\begin{array}{c}
    \omega|_\Sigma=\frac12(\partial_\mu B_\nu-\partial_\nu B_\mu +\pi^{\lambda\gamma}\partial_\mu B_\lambda \partial_\nu B_\gamma)\dd x^\mu\wedge \dd x^\nu =0\,.
    \end{array}
\end{equation}
Unlike the previous example, this equation is nonlinear.  However, each Lagrangian bisection lying infinitely close to the base  $X$ is still locally represented as $B_\mu=\partial_\mu\varepsilon(x)$ for an infinitesimal  $\varepsilon(x)$. Substituting this in (\ref{AB}) gives the expected gauge transformation  $\delta_\varepsilon A_\mu=\partial_\mu\varepsilon+\{A_\mu,\varepsilon\}$, cf. (\ref{PED}).

\end{example}

\begin{example}\label{Exam3}({\it Lie--Poisson structures.}) Suppose $X$ is diffeomorphic to a Cartesian space with coordinates $x^\mu$. Then it is legitimate to consider linear Poisson brackets of the form
\begin{equation}\label{LP}
    \{x^\mu,x^\nu\}=f^{\mu\nu}_\lambda x^\lambda\,.
\end{equation}
It follows from the Jacobi identity that  $f^{\mu\nu}_\lambda$  are structure constants of some Lie algebra $\mathfrak{g}$ with the commutation relations $[e^\mu,e^\nu]=f^{\mu\nu}_\lambda e^\lambda$. This allows us to identify $X$ with the dual space of the Lie algebra $\mathfrak{g}^\ast$. Let $G$ denote a Lie group with the Lie algebra $\mathfrak{g}$. As a symplectic groupoid integrating the Lie--Poisson structure (\ref{LP}) one can take the cotangent bundle $T^\ast G$ endowed with the canonical symplectic structure.  Since the cotangent bundle of any Lie group is trivial, one has a trivializing diffeomorphism  $T^\ast G\simeq G\times \mathfrak{g}^\ast$; the points of $G\times \mathfrak{g}^\ast$ are the pairs $(g,x)$ with $g\in G$ and $x\in \mathfrak{g}^\ast$.  
In coordinates adapted to the trivialization, the Liouville $1$-form on $T^\ast G$ reads 
\begin{equation}
    \vartheta=\langle x, g^{-1}\dd g\rangle\,.
\end{equation}
Here $g^{-1}\dd g$ is a left-invariant $1$-form on $G$ with values in $\mathfrak{g}$, and the triangle brackets stand for the natural pairing between the spaces $\mathfrak{g}^\ast$ and $\mathfrak{g}$.  Expanding over the basis above, we can write $g^{-1}\dd g=\gamma_\mu e^\mu$, where $\{\gamma_\mu\}$ is a basis of left-invariant $1$-forms on $G$. Hence, $\vartheta = \langle x, g^{-1}\dd g\rangle  =x^\mu \gamma_{\mu}$. This leads to the symplectic $2$-form 
\begin{equation}
\begin{array}{c}
    \omega=\dd \vartheta=dx^\mu\wedge \gamma_\mu-\frac12 x^\lambda f^{\mu\nu}_\lambda\gamma_\mu\wedge \gamma_\nu\,.
    \end{array}
\end{equation}
The source and target maps are given by
\begin{equation}
    \sfs (g,x)=x\,,\qquad \sft (g,x)= \mathsf{Ad}_g^\ast (x)\,,
\end{equation}
where $\mathsf{Ad}^\ast$ denotes the coadjoint representation of $G$. The group of bisections consists of graphs $\Sigma =(\sigma(x), x)\subset G\times \mathfrak{g}^\ast$ of smooth mappings $\sigma:  \mathfrak{g}^\ast\rightarrow G$. If $\Sigma' =(\sigma'(x), x)$
is another bisection of $\mathscr{B}(T^\ast G)$, then 
\begin{equation}\label{complaw}
    \Sigma \Sigma' =\big(\sigma(\mathsf{Ad}^\ast_{\sigma'(x)}(x))\sigma'(x), x\big) \,.
\end{equation}
It is straightforward to check that both $\vartheta$ and $\omega$ enjoy multiplicativity (\ref{mult}), see Eq. (\ref{tss}) below.  

\end{example}

According to postulate P3, the electromagnetic field on the Poisson manifold $(X,\pi)$ is described by bisections $\Sigma$ of the symplectic groupoid $\mathcal{G}\rightrightarrows X$. By postulate P4, two field configurations $\Sigma_1$ and $\Sigma_2$ are considered gauge equivalent, $\Sigma_1\sim \Sigma_2$, if there is a Lagrangian bisection $\Sigma$ such that $\Sigma_1=\Sigma_2\Sigma$. 
This allows us to identify the classes of gauge equivalent electromagnetic potentials with the elements of the right quotient $\mathscr{B}(\mathcal{G})/\mathscr{L}(\mathcal{G})$. Thus, the Lagrangian bisections should be regarded as `pure gauge' field configurations. Now, we would like to introduce an object that would measure the deviation of a bisection from being Lagrangian.  In ordinary electrodynamics, such a role is played by the strength tensor $F=\dd A$. 
The definition of a Lagrangian bisection suggests two natural generalizations  of the usual strength tensor \cite{kupriyanov2023symplectic}:
\begin{equation}\label{sFtF}
    F^\sfs(\Sigma):=\Sigma_\sfs^\ast\omega \quad\text{and}\quad  F^\sft(\Sigma):=\Sigma_\sft^\ast\omega \,.
\end{equation}
Both quantities are given by closed $2$-forms on $X$, which vanish on Lagrangian bisections $\Sigma\in \mathscr{L}(\mathcal{G})$.  
The multiplicative properties (\ref{mult}) are equivalent to the identities
\begin{equation}\label{mult1}
F^\sfs(\Sigma_1\Sigma_2)= F^\sfs(\Sigma_2)+l^\ast_{\Sigma_2}F^\sfs(\Sigma_1)\,,\qquad F^\sft(\Sigma_1\Sigma_2)= F^\sft(\Sigma_1)+r^\ast_{\Sigma_1}F^\sft(\Sigma_2)\,.
\end{equation}
These  generalize the additivity of the ordinary strength tensor\footnote{As explained in Example \ref{Exam1}, for the zero Poisson bracket $l_\Sigma=r_\Sigma=\mathrm{id}_X$.}, $F(A_1+A_2)=F(A_1)+F(A_2)$.
Applying these formulas to $\Sigma\in \mathscr{B}(\mathcal{G})$ and $\Sigma'\in \mathscr{L}(\mathcal{G})$, we obtain  the following behaviour of the strength tensors (\ref{sFtF})  under the gauge transformations $\Sigma\mapsto \Sigma \Sigma'$:
\begin{equation}
    F^\sfs(\Sigma\Sigma')= l^\ast_{\Sigma'}F^\sfs(\Sigma)\,,\qquad F^\sft(\Sigma\Sigma')= F^\sft(\Sigma) \,.
\end{equation}
($F^\sfs(\Sigma')=F^\sft(\Sigma')=0$, since $\Sigma'$ is a Lagrangian bisection.) Like the usual strength tensor $F=\dd A$, the $2$-form $F^\sft(\Sigma)$ is gauge invariant, while $F^\sfs(\Sigma)$ is only covariant under the  action of the gauge group $\mathscr{L}(\mathcal{G})$. The covariant and invariant strength tensors are related to each other by  field-dependent diffeomorphisms, 
\begin{equation}\label{FeF}
    F^\sfs(\Sigma)=l_\Sigma^\ast F^\sft(\Sigma)\,,\qquad    F^\sft(\Sigma)=r_\Sigma^\ast F^\sfs(\Sigma)\,.
\end{equation}
Notice that if, under a suitable parametrization of $\Sigma$, the tensor $F^\sfs(\Sigma)$ is local with respect to the gauge potential, then $F^\sft(\Sigma)$ is nonlocal, and vice versa. 

One can use the tensors (\ref{sFtF}) to construct gauge invariant Lagrangians and field equations. As usual, this requires the choice of 
a background metric $g$. The Lagrangian is  given by a top-form 
\begin{equation}\label{Linv}
    \mathcal{L}_{\mathrm{inv}}=\mathcal{L}(F^\sft, g,\pi,\ldots)\in \sfLambda^{\mathrm{top}}(X)
\end{equation}
constructed, presumably in a local manner, from the strength tensor $F^\sft$, metric field $g$,  Poisson bivector $\pi$, and other background fields, if any. By construction, the integral
\begin{equation}
    S_{\mathrm{em}}[\Sigma]=\int_X  \mathcal{L}(F^\sft, g,\pi,\ldots)
    \end{equation}
defines a gauge invariant action functional.  We also set
\begin{equation}\label{Lcov}
    \mathcal{L}_{\mathrm{cov}}=l^\ast_\Sigma\mathcal{L}_{\mathrm{inv}}=\mathcal{L}(F^\sfs, g^\Sigma, \pi^\Sigma,\ldots)\,.
\end{equation}
Here we introduced the following notation: if $T$ is a tensor field on $X$, then $T^\Sigma=l_\Sigma^\ast T$ is the result of applying to $T$ the diffeomorphism $l_\Sigma: X\rightarrow X$. Unlike $\mathcal{L}_{\mathrm{inv}}$, the Lagrangian $\mathcal{L}_{\mathrm{cov}}$ is not gauge invariant, but transforms covariantly under the gauge transformations. With the help of the identity $l^\ast_{\Sigma\Sigma'}=l^\ast_{\Sigma'}l^\ast_{\Sigma}$, we find
\begin{equation}
    \Sigma\rightarrow \Sigma\Sigma'\quad \Rightarrow\quad \mathcal{L}_{\mathrm{cov}}\rightarrow l_{\Sigma'}^\ast\mathcal{L}_{\mathrm{cov}}\qquad \forall \Sigma'\in \mathscr{L}(\mathcal{G})\,.
\end{equation}
Relation (\ref{Lcov}) means that both Lagrangians give the same (gauge invariant) action functional,
\begin{equation}
    \int_X\mathcal{L}_{\mathrm{cov}}= S_{\mathrm{em}}[\Sigma] =\int_X\mathcal{L}_{\mathrm{inv}}\,.
\end{equation}

\begin{example} Consider an $n$-dimensional Minkowski space with metric $\eta_{\mu\nu}$ and a constant Poisson bivector $\pi^{\mu\nu}$, as in Example \ref{Exam2}. Each bisection $\Sigma$ defines and is defined by a map 
\begin{equation}
    \Sigma_\sfs(x)=(A_\mu(x), x^\nu)\subset \mathbb{R}^n\times \mathbb{R}^n\,.
\end{equation}
It follows from the definition (\ref{E2ST}) that $\sfs\circ \Sigma_\sfs=\mathrm{id}$ as it must. The pullback of the symplectic $2$-form (\ref{E2SS}) with respect to $\Sigma_\sfs$ then gives the gauge covariant strength tensor. 
\begin{equation}\label{EFcov}
\begin{array}{c}
    F^\sfs (A)=\Sigma_\sfs^\ast(\omega)= \frac12(\partial_\mu A_\nu-\partial_\nu A_\mu +\pi^{\lambda\gamma}\partial_\mu A_\lambda \partial_\nu A_\gamma)\dd x^\mu\wedge \dd x^\nu \,,
    \end{array}
\end{equation}
which is a $2$-form on Minkowski space. As seen, the strength tensor depends on the first derivatives of the gauge potential $A=A_\mu(x)\dd x^\mu$. Similarly, to construct a gauge invariant  strength tensor $F^\sft(A)$ we use the map 
$
    \Sigma_\sft(x)=(A_\mu(\tilde x), \tilde x^\nu)
$,
where the point $\tilde x$ is determined from the condition $\sft\circ \Sigma_\sft=\mathrm{id}$. With (\ref{E2ST}) we obtain the equation
\begin{equation}\label{xtx}
    x^\mu=\tilde x^\mu+\pi^{\mu\nu}A_\nu(\tilde x)\,,
\end{equation}
which defines an implicit function $\tilde x^\mu=\tilde x^\mu(x)$. We have
\begin{equation}\label{Fxxx}
    F^\sft (A)=\Sigma_\sft^\ast(\omega)= \frac12(\partial_\mu A_\nu-\partial_\nu A_\mu +\pi^{\lambda\gamma}\partial_\mu A_\lambda \partial_\nu A_\gamma)\dd \tilde x^\mu\wedge \dd \tilde x^\nu|_{\tilde x=\tilde x(x)}\,.
\end{equation}
Eq. (\ref{xtx}) defines a diffeomorphism $l_\Sigma: \mathbb{R}^n\rightarrow \mathbb{R}^n$ of Minkowski space generated by the bisection $\Sigma=(A_\mu(x), x^\nu)$. The inverse diffeomorphism $r_\Sigma=l_\Sigma^{-1}$ corresponds to the substitution ${\tilde x=\tilde x(x)}$ in (\ref{Fxxx}). Hence, 
$F^\sft=r^\ast_\Sigma F^\sfs$ in accordance with general formula (\ref{FeF}). In contrast to the gauge covariant tensor (\ref{EFcov}), the  invariant tensor (\ref{Fxxx}) is highly nonlocal and nonlinear.

To apply the above prescription for constructing a gauge covariant Lagrangian (\ref{Lcov}), we also need the formula for the Minkowski metric twisted by the diffeomorphism $l_\Sigma$. We find
\begin{equation}
    \eta^\Sigma=l_\Sigma^\ast \eta=\eta_{\alpha\beta}\,\big(\delta^\alpha_\mu+\pi^{\alpha\lambda}\,\partial_\mu A_\lambda\big)\,\big(\delta^\beta_\nu+\pi^{\beta\gamma}\partial_\nu A_\gamma\big)\dd x^\mu\dd x^\nu\,.
    \end{equation}
Like the covariant strength tensor, the twisted metric depends on the first derivatives of the gauge potential. With $F^\sfs$ and $\eta^\Sigma$ one can construct a large number of gauge invariant action functionals satisfying the correspondence principle. In\cite{kupriyanov2023symplectic}, the following Lagrangian was considered: 
\begin{equation}\label{ELcov}
    \mathcal{L}_{\mathrm{cov}}=\eta_\Sigma^{\mu \lambda}\,\eta_\Sigma^{\nu \gamma}\,\big(F^\sfs_{\mu\nu}+ F^\sfs_{\mu \alpha}\,\pi^{\alpha\beta}\,F^\sfs_{\beta\nu}\big)\,\big(F^\sfs_{\lambda\gamma}+ F^\sfs_{\lambda \sigma}\,\pi^{\sigma\delta}\,F^\sfs_{\delta\gamma}\big)\,\dd^nx= \eta^{\mu\nu}\,\eta^{\lambda \gamma}\, F_{\mu \lambda}\, F_{\nu \gamma} \, \dd^nx \,.\end{equation}
Here $\eta^{\mu\nu}_\Sigma$ is the inverse matrix to $\eta^\Sigma_{\mu\nu}$ and 
$
  F_{\mu\nu}=\partial_\mu A_\nu-\partial_\nu A_\mu+\{A_\mu,A_\nu\}  
$. 
One can easily verify that under the standard gauge transformations $\delta_\varepsilon A_\mu=\partial_\mu\varepsilon+\{A_\mu, \varepsilon\}$ generated by infinitesimal Lagrangian bisections $\Sigma'=(\dd \varepsilon, x)$, the components of the matrix $F_{ab}$ transform as $\delta_\varepsilon F_{\mu\nu}=\{\varepsilon, F_{\mu\nu}\}$. Together with the invariance of the canonical volume form\footnote{Another option is to take the gauge covariant volume form $\sqrt{|\det(\eta_{\mu\nu}^\Sigma)|}\,\dd^n x$. 
Unlike $\dd^n x$, this volume form depends on the gauge field.
} $\dd^n x$ under the Hamiltonian flow $V_{\varepsilon}=\{\varepsilon, -\}$ this ensures the gauge covariance of the Lagrangian (\ref{ELcov}). 
In the commutative limit $\pi=0$, (\ref{ELcov}) goes over into the standard Lagrangian of the electromagnetic field. 
\end{example}

For other examples of gauge covariant Lagrangians  associated with Lie--Poisson structures we refer the reader to \cite{Kupriyanov:2022ohu, Kupriyanov:2023gjj}.

\section{Minimal coupling to matter fields}\label{S3}

The invariant and covariant Lagrangians (\ref{Linv}, \ref{Lcov}) represent a general recipe to introduce a nonminimal coupling of the electromagnetic field to arbitrary external fields. Now, we turn to the issue of {\it minimal} interaction with a proper commutative limit, a problem discussed at length in $\S$\ref{S1}. Let $\Phi$ denote a complex field on $X$, which may well be multicomponent, although 
we suppress all possible indices. 
The gauge group $\mathscr{L}(\mathcal{G})$ of Poisson electrodynamics is assumed to act on $\Phi$ in a pure algebraic way 
\begin{equation}\label{gtr1}
    \Phi\quad\rightarrow \quad\mathcal{U}(\Sigma_1)\Phi= e^{-i\alpha (\Sigma; \Sigma_1)} \Phi\,.
\end{equation}
Here the bisection $\Sigma$ describes the electromagnetic field and  $\Sigma_1$ is a Lagrangian bisection determining the gauge transformation.
The phase factor $e^{-i\alpha(\Sigma; \Sigma_1)}$ should be chosen to define a representation of the gauge group $\mathscr{L}(\mathcal{G})$ in the space of fields $\Sigma$ and $\Phi$, i.e., $\mathcal{U}(\Sigma_1)\mathcal{U}(\Sigma_2)=\mathcal{U}(\Sigma_1\Sigma_2)$. This is equivalent to the relation 
\begin{equation}\label{sss}
   (\delta\alpha)(\Sigma;\Sigma_1,\Sigma_2):=\alpha(\Sigma\Sigma_1;\Sigma_2)-\alpha(\Sigma;\Sigma_1\Sigma_2) +\alpha(\Sigma;\Sigma_1)=0\quad (\,\mathrm{mod} \;2\pi\,)\,,
    \end{equation}
    $$
    \forall\, \Sigma\in\mathscr{B}(\mathcal{G})\,,\quad \forall\, \Sigma_1,\Sigma_2\in\mathscr{L}(\mathcal{G})\,.    
    $$
In this equation, one can recognize the cocycle condition for a $1$-cocycle of the group $\mathscr{L}(\mathcal{G})$. 
The general definition of group cohomology is given in Appendix \ref{App}. 

Evaluating Rel. (\ref{sss}) on the bisections $\Sigma_1$ and $\Sigma_2$ that are infinitely close to $X$, one can deduce the cocycle condition (\ref{1c}) for the corresponding Lie algebra $\mathcal{L}$. Let us look at this point in more detail. It is well known that the Lie algebra $\mathcal{L}$ of the group $\mathscr{L}(\mathcal{G})$ coincides with the space $\mathsf{Z\Lambda}^1(X)$ of closed $1$-forms on $X$ endowed with the Lie bracket\footnote{This bracket  makes the cotangent bundle $T^\ast X$ into the Lie algebroid with anchor $\pi: T^\ast X\rightarrow TX$. In this paper, however, we will not enlarge on the Lie algebroid interpretation of the infinitesimal gauge transformations. }
\begin{equation}\label{BB}
    [\gamma_1,\gamma_2]:=\dd \pi(\gamma_1, \gamma_2) \qquad \forall \gamma_1, \gamma_2 \in \mathsf{Z\Lambda}^1(X)\,,
\end{equation}
see e.g. \cite[Thm. 4.5]{Rybicki2001OnTG}. For a simply connected $X$, the first cohomology group ${H}^1(X)=0$ and the de Rham differential $\dd: \mathsf{\Lambda}^0(X)\rightarrow \mathsf{\Lambda}^1(X)$ defines an isomorphism $\mathsf{Z\Lambda}^1(X)\simeq \mathsf{\Lambda}^0(X)/\mathbb{R}$. In this case, each closed $1$-form is exact and we can put $\gamma=\dd \varepsilon$ for some $\varepsilon\in\mathsf{\Lambda}^0(X)$. The Lie algebra (\ref{BB}) of gauge transformations is then isomorphic to the Poisson-bracket algebra  of functions $\mathsf{\Lambda}^0(X)$ factored by the centre $\mathbb{R}$. In other words, the Lie algebra of smooth functions on $X$ with respect to the Poisson bracket is a central extension of the Lie algebra of gauge transformations (\ref{BB}).

As usual, the minimal coupling is defined through a covariant derivative that respects the gauge transformations (\ref{gtr1}). The form of the gauge transformations suggests to look for such a covariant derivative in the form 
\begin{equation}\label{covder1}
    D\Phi=\dd \Phi +i\theta(\Sigma) \Phi\,,
\end{equation}
where the $1$-form $\theta(\Sigma)\in \mathsf{\Lambda}^1(X)$ is a composite field built from the electromagnetic potentials. The covariance condition $D\mathcal{U}(\Sigma_1) \Phi=\mathcal{U}(\Sigma_1) D\Phi$ implies that 
\begin{equation}\label{da}
    (\delta\theta)(\Sigma;\Sigma_1)=\dd \alpha(\Sigma;\Sigma_1)   \,,
    \end{equation}
where 
\begin{equation}
    (\delta\theta)(\Sigma;\Sigma_1):=\theta(\Sigma\Sigma_1)-\theta(\Sigma)\,.
\end{equation}
Applying the coboundary operator $\delta$ to both sides of Eq. (\ref{da}), we get $\dd \delta\alpha=0$. Hence, one can think of the cocycle condition (\ref{sss}) as a compatibility condition for the linear nonhomogeneous equation (\ref{da}).  On the other hand, acting on (\ref{da}) by the exterior differential $\dd$, we find that the $2$-form 
$\psi=\dd \theta (\Sigma)$ defines a $0$-cocyle of the group $\mathscr{L}(\mathcal{G})$, i.e.,
\begin{equation}
    (\delta\psi)(\Sigma;\Sigma_1):=\psi(\Sigma\Sigma_1)-\psi(\Sigma)=0\,.
\end{equation}
Translated  into physical language, this cocycle condition states that $\psi(\Sigma)$ is a gauge invariant $2$-form. Notice that, by construction, $\dd\psi=0$. 

Now, to find a solution to equations (\ref{sss}), we can move backwards starting from a $\dd$- and $\delta$-closed $2$-form $\psi(\Sigma)$.  Let us make a couple of simplifying technical assumptions about the topology of $X$, namely, 
\begin{equation}\label{TR}
    {H}^0(X)=\mathbb{R} \,,\qquad {H}^1(X)={H}^2(X)=0\,.
\end{equation} Then there is a $1$-form $\theta(\Sigma)$ such that $\psi=\dd\theta$. Since the operators  $\dd$ and $\delta$ commute, the last equality implies that $\dd\delta \theta=0$. Hence, there exists a $0$-form $\alpha$ such that $\delta\theta =\dd\alpha$. Applying the operator $\delta$ once again, we get $\dd \delta\alpha=0$, that is, $\delta\alpha=\phi$ where $\phi=\phi(\Sigma;\Sigma_1,\Sigma_2)$ is a constant function on $X$.  By construction, $\delta \phi=0$.  The transition from $\psi$ to $\phi$ involves an arbitrary choice. For example, we can add to $\alpha$ an arbitrary constant function $\kappa(\Sigma;\Sigma_1)$. This results in the shift $\phi\rightarrow \phi+\delta \kappa$. By construction, $\phi=\delta\alpha$ is a trivial $2$-cocycle with values in functions. However, it may well be nontrivial if we restrict ourselves to the subcomplex  ${C}(\mathcal{G}, \mathbb{R}) \subset{C}(\mathcal{G}, \mathsf{\Lambda}^0(X))$ of constant functions on $X$. Triviality of the cocycle $\phi$ as an element of 
${C}^2(\mathcal{G}, \mathbb{R})$ would imply the existence of a $1$-cochain $\kappa\in {C}^1(\mathcal{G}, \mathbb{R})$ such that $\phi=\delta \kappa$. It follows from the discussion above that every $d$- and  $\delta$-closed $2$-form  $\psi(\Sigma)$ gives rise to a well defined cohomology class $[\phi]\in H^2(\mathcal{G}, \mathbb{R})$.  If $[\phi]=0$, that is, $\phi=\delta \kappa$ for some $\kappa\in C^1(\mathcal{G},\mathbb{R})$, then $\theta$ and $\alpha'=\alpha- \kappa$  provide a solution to Eqs. (\ref{da}) and (\ref{sss}).

It is also interesting to see what happens if $[\phi]\neq 0$. In this case, one cannot shift $\phi$ by a constant $1$-cochain $\kappa$ 
to make it into a $1$-cocycle. Exponentiating the equation $\delta\alpha=\phi$, we find that the functions $\mathcal{U}(\Sigma;\Sigma_1)=e^{-i\alpha(\Sigma;\Sigma_1)}$ satisfy the relation
\begin{equation}\label{Prep}
    \mathcal{U}(\Sigma;\Sigma_1)\mathcal{U}(\Sigma\Sigma_1;\Sigma_2)=e^{-i\phi(\Sigma;\Sigma_1,\Sigma_2)}\mathcal{U}(\Sigma;\Sigma_1\Sigma_2)\,.
\end{equation}
This is the defining condition of the so-called {\it projective representation} of the group $\mathscr{L}(\mathcal{G})$. Of course, it may happen that  $\phi(\Sigma; \Sigma_1,\Sigma_2)=2\pi n$ for some integer $n$, in which case the phase factor disappears. The adjective {\it projective} means that the $\mathcal{U}$'s define a genuine representation in the space of complex-valued functions $\Phi$ considered modulo constant phase factors,  $\Phi\sim e^{-i\phi}\Phi$. Such an identification is a common thing in quantum mechanics, when one treats the complex field $\Phi$ as a wave function representing a quantum state. Another possible interpretation of  (\ref{Prep}) is by means  of a central extension of the group $\mathscr{L}(\mathcal{G})$, see \cite[Ch. 13]{raczka1986theory}. We will not dwell on relation (\ref{Prep}) in more detail, since it apparently does not occur in the problem under consideration.

Turning back to our problem, we can take as the double cocycle $\psi$ the invariant strength tensor $F^\mathsf{t}(\Sigma)$. Under assumptions (\ref{TR}), this yields a $1$-form $\theta(\Sigma)$ satisfying the equation $\dd \theta=F^{\mathsf{t}}$ as well as a $1$-cochain $\alpha(\Sigma;\Sigma_1)$ with the property that $\delta\theta=\dd \alpha$. This data is enough to define the gauge transformation (\ref{gtr1}) and covariant derivative (\ref{covder1}).

One can take a more general point of view and treat $\Phi$ as a section of a complex line bundle $\mathbb{L}$ over $X$. Then $\theta$ becomes a connection $1$-form in $\mathbb{L}$ with curvature $F^{\mathsf{t}}$. As is well known, such an interpretation is only possible if the integral of $F^\mathsf{t}$ over any  closed oriented $2$-surface $S\subset X$ is an integral multiple of  $2\pi$, see e.g. \cite[Ch. 8.3]{woodhouse1992geometric}. Furthermore, if $X$ is simply connected and the integrality condition holds, then the line bundle $\mathbb{L}$  is uniquely determined by the invariant curvature $F^\mathsf{t}$. Now, we can argue that the integrality condition is automatically satisfied for all bisections $\Sigma$ that can be continuously  deformed to the base $X$ over $\mathcal{G}$. More precisely, we assume  the existence of a smooth map $\sigma: X\times [0,1]\rightarrow \mathcal{G}$ such that $\sigma(X, 0)=\Sigma$ and $\sigma(X, 1)=X$. (In this case, one says that the submanifolds $\Sigma$ and $X$ are homotopy equivalent.) For example, this happens for bisections that are sufficiently close to $X$ 
(weak electromagnetic fields) or bisections belonging to one-parameter subgroups of $\mathscr{B}(\mathcal{G})$. Indeed, for any closed oriented surface $S\subset X$, $\Sigma_\sft(S)\subset \Sigma$. The homotopy equivalence of $\Sigma$ and $X$ implies that $\Sigma_\sft(S)$ is homotopy equivalent to some surface $S'\subset X$. Since $\omega|_X=0$, the statement follows from the chain of equalities
\begin{equation}
    \int_SF^{\mathsf{t}}=\int_S\Sigma_{\mathsf{t}}^\ast \omega=\int_{\Sigma_\mathsf{t}(S)}\omega=\int_{S'}\omega+\int_{\partial  C}\omega=\int_{ C }\dd \omega=0\,,
\end{equation}
where $C=\sigma(S',[0,1])\subset \mathcal{G}$ and we used Stokes' theorem. Therefore, we can relax the topological restrictions (\ref{TR}) by replacing them with the integrality condition on the strength tensor $F^\sft(\Sigma)$, which is automatically satisfied for weak electromagnetic fields $\Sigma$.   A more detailed discussion of topological aspects of Poisson electrodynamics will be given elsewhere. We conclude this paper with a couple of examples illustrating the general formalism. 

\begin{example} We continue with our through example of a constant Poisson bracket. Both covariant (\ref{EFcov}) and invariant (\ref{Fxxx}) strength tensors of the electromagnetic field $A=A_\mu(x)\dd x^\mu$ are  exact $2$-forms:
\begin{equation}\label{FF}
\begin{array}{l}
F^\sfs=\Sigma_\sfs^\ast(\omega)=\dd \big(A_\mu\dd x^\mu+\frac12\pi^{\mu\nu}A_\mu\partial_\lambda A_\nu \dd x^\lambda\big)\,,\\[3mm]
    F^\sft=r_\Sigma^\ast F^{\sfs}=\dd r_\Sigma^\ast\big(A_\mu\dd x^\mu+\frac12\pi^{\mu\nu}A_\mu\partial_\lambda A_\nu \dd x^\lambda\big)=\dd \theta\,.
    \end{array}
\end{equation}
Unfortunately, there is no closed expression for the diffeomorphism $r_\Sigma: \mathbb{R}^n\rightarrow \mathbb{R}^n$ generated by the bisection $\Sigma=(A_\mu(x),x^\nu)$. By iterating Eq. (\ref{xtx}) for the inverse diffeomorphism $l_\Sigma$, we can construct $r_\Sigma(x)$ as an expansion in powers of the gauge field $A$ or,  equivalently, the Poisson bivector $\pi$. The expansion starts as
\begin{equation}\label{rexp}
    r_\Sigma(x)^\mu=x^\mu-\pi^{\mu\nu}A_\nu(x)+\mathcal{O}(\pi^2)\,.
\end{equation}
Since $X=\mathbb{R}^n$ obviously satisfies the topological conditions (\ref{TR}), one can regard the $2$-form $F^\sft$ as the curvature of a trivial line bundle  $\mathbb{L}\simeq \mathbb{R}^n\times \mathbb{C}$. 
On substituting (\ref{rexp}) into (\ref{FF}), we obtain the expansion for the corresponding connection   
\begin{equation}
\begin{array}{c}
    \theta(A)=A_\mu\dd x^\mu+\pi^{\mu\nu}A_\mu\big(\partial_\nu A_\lambda-\frac12\partial_\lambda A_\nu\big)\dd x^\lambda+\mathcal{O}(\pi^2)\,.
    \end{array}
\end{equation}
By definition, the curvature  $F^\sft(A)$ is invariant under the gauge transformation $\delta_\varepsilon A=\dd \varepsilon +\{A,\varepsilon\}$, that is, 
$\delta_\varepsilon F^\sft=\dd \delta_\varepsilon\theta=0$. This means that the gauge variation of $\mathcal{\theta}(A)$ must be a $\dd$-exact $1$-form. A straightforward computation gives 
\begin{equation}
\begin{array}{c}
    \delta_\varepsilon \theta(A)=\dd \alpha (A,\varepsilon)\,,\qquad \alpha(A,\varepsilon)= \varepsilon +\frac12\pi^{\mu\nu}A_\mu\partial_\nu \varepsilon +\mathcal{O}(\pi^2)\,.
    \end{array}
\end{equation}
The function $\alpha(A,\varepsilon)$ generates the gauge transformation
\begin{equation}
    \delta_\varepsilon\Phi=-i\alpha(A,\varepsilon)\cdot \Phi
\end{equation}
 of the complex field $\Phi$, cf. (\ref{ans}). If $\mathcal{L}(\Phi,\partial_\mu \Phi)$ is a $U(1)$-invariant Lagrangian of the complex field $\Phi$, then the minimal coupling to the electromagnetic field $A$ is introduced by the standard replacement of the partial derives of $\Phi$ with the covariant derivatives $D_\mu=\partial_\mu+i\theta_\mu(A)$. 
\end{example}

\begin{example} Consider Lie--Poisson structures from Example \ref{Exam3}. Each bisection is defined by a smooth map $\sigma: \mathfrak{g}^\ast\rightarrow G$. Since the corresponding symplectic form is exact, $\omega=\dd \vartheta$, we can write
\begin{equation}\label{FFc}
\begin{array}{rcl}
F^\sfs&=&\Sigma_\sfs^\ast(\omega)=\dd \langle x,  \sigma(x)^{-1}\dd \sigma(x)\rangle \,,\\[3mm]
F^\sft&=&r_\Sigma^\ast F^{\sfs}=\dd r_\Sigma^\ast
\langle x,  \sigma(x)^{-1}\dd \sigma(x)\rangle =
\dd \theta(\Sigma)\,.
    \end{array}
\end{equation}
The diffeomorphism $r_\Sigma: \mathfrak{g}^\ast\rightarrow  \mathfrak{g}^\ast$ associated with the bisection $\Sigma=(\sigma(x),x)\subset G\times \mathfrak{g}^\ast$ is inverse to the mapping\footnote{For bisections close to the base $\mathfrak{g}^\ast$ -- the identity element of the group $\mathscr{B}(T^\ast G)$ -- one can approximatly invert relation (\ref{map}) as $\tilde x\approx \mathsf{Ad}^\ast_{\sigma(x)^{-1}}(x)$. }
\begin{equation}\label{map}
  \tilde x\mapsto {x}=l_\Sigma(\tilde x)=\mathsf{Ad}_{\sigma(\tilde x)}^\ast(\tilde x)\,.
\end{equation}
It follows from (\ref{FFc}) that
\begin{equation}\label{teta}
     \theta(\Sigma)=\Sigma_{\sft}^\ast\vartheta=r_\Sigma^\ast
\langle x,  \sigma(x)^{-1}\dd \sigma(x)\rangle=\langle \tilde x,  \sigma^{-1}(\tilde x)\dd \sigma(\tilde x)\rangle|_{\tilde{x}=\tilde{x}(x)}\,.\end{equation}
 If $\Sigma'=(\sigma'(x), x)$ is a Lagrangian bisection, then
\begin{equation}
    \omega|_{\Sigma'}= \dd \langle x,\sigma'(x)^{-1}\dd \sigma'(x)\rangle =0\,.
\end{equation}
Hence, there exists a smooth function $\varphi(\Sigma')\in \sfLambda^0(\mathfrak{g}^\ast)$ such that 
\begin{equation}\label{fi}
    (\Sigma'_\sfs)^\ast\vartheta=\langle x,\sigma'(x)^{-1}\dd \sigma'(x)\rangle =\dd \varphi(\Sigma')\,.
    \end{equation}
With the composition law (\ref{complaw}), we find
\begin{equation}\label{tss}
\begin{array}{rcl}
    \theta(\Sigma\Sigma')&=&r^\ast_{\Sigma\Sigma'}\big\langle x,  \sigma'(x)^{-1}\sigma(\mathsf{Ad}^\ast_{\sigma'(x)}(x))^{-1}\dd \big(\sigma(\mathsf{Ad}^\ast_{\sigma'(x)}(x))\sigma'(x)\big)\big\rangle\\[5mm]
    &=&r^\ast_{\Sigma\Sigma'}\langle x,  \sigma'(x)^{-1}\dd \sigma'(x)\rangle +r^\ast_{\Sigma\Sigma'}\big\langle x,  \sigma'(x)^{-1}\big[\sigma(\mathsf{Ad}^\ast_{\sigma'(x)}(x))^{-1}\dd \sigma(\mathsf{Ad}^\ast_{\sigma'(x)}(x)\big)\big]\sigma'(x)\big\rangle    
    \\[5mm]
    
                         &=&r^\ast_{\Sigma}r^\ast_{\Sigma'}\langle x,  \sigma'(x)^{-1}\dd \sigma'(x)\rangle  +r^\ast_{\Sigma}r^\ast_{\Sigma'}\big\langle  \mathsf{Ad}^\ast_{\sigma'(x)}(x),  \sigma(\mathsf{Ad}^\ast_{\sigma'(x)}(x))^{-1}\dd \sigma(\mathsf{Ad}^\ast_{\sigma'(x)}(x))\big\rangle\\[5mm]   
                         &=&r^\ast_\Sigma \theta(\Sigma')+\theta(\Sigma)= \theta(\Sigma) +\dd r^\ast_{\Sigma\Sigma'} \varphi(\Sigma')  \,. \end{array}
    \end{equation}
    Here we used Eqs. (\ref{teta}), (\ref{fi}) and the definition of the coadjoint representations $\langle \mathsf{Ad}^\ast_g(x),y\rangle:=\langle x, \mathsf{Ad}_{g^{-1}}(y)\rangle$. Among other things, we have proved that the symplectic potential $\vartheta$ is multiplicative, cf. (\ref{mult}).   Setting $\alpha(\Sigma; \Sigma')=r^\ast_{\Sigma\Sigma'} \varphi(\Sigma') $, we define finite gauge transformations as $\Sigma\mapsto \Sigma\Sigma'$ and $\Phi\mapsto e^{-i\alpha(\Sigma;\Sigma')}\Phi$ for any Lagrangian bisection $\Sigma'\in \mathscr{L}({T^\ast G})$. The minimal coupling is now introduced through the covariant differential $D\Phi=\dd \Phi+i\theta(\Sigma)\Phi$.
\end{example}

As a final remark, we note that the same $1$-form $\theta(\Sigma)$ allows one to introduce a minimal coupling of a point charged particle to an external electromagnetic field $\Sigma$. This is achieved by adding the standard term $e\theta_\mu\dot x^\mu$ to the Lagrangian of the particle. The gauge invariance of the interaction is obvious.

\section{Conclusion}

To summarized, we have given a solution to a long-standing problem of minimal interaction in Poisson electrodynamics. Until now, such an interaction was known only for the matter fields in the adjoint representation of the gauge group. This, however, did not reproduce the ordinary electrodynamics in the commutative limit. A key geometric ingredient of our approach is the interpretation of electromagnetic field configurations as bisections of a symplectic groupoid integrating a given Poisson structure. Using the geometry of symplectic groupoids, we construct the gauge covariant and invariant strength tensors and use the latter to define a covariant derivative that implements  minimal interaction with a proper commutative limit. If $\mathcal{L}(\Phi, \partial_\mu \Phi, \ldots )$ is a $U(1)$ invariant Lagrangian for complex matter fields $\Phi$, which may also depend on some other dynamical or background fields, then the minimal electromagnetic coupling is described by the gauge invariant Lagrangian $\mathcal{L}_{\mathrm{inv}}=\mathcal{L}(\Phi, D_\mu\Phi, \ldots)$. The covariant differential $D=\dd+i\theta(\Sigma)$ is determined by a connection $1$-form $\theta(\Sigma)$, which is now a nonlinear and nonlocal function of a bisection $\Sigma$. As with the Lagrangian of a pure electromagnetic field (\ref{Linv}, \ref{Lcov}), one can eliminate nonlocality by passing to a gauge-covariant Lagrangian. 
\begin{equation}\label{tD}
\mathcal{L}_{\mathrm{cov}}=l^\ast_\Sigma\mathcal{L}_{\mathrm{inv}}=\mathcal{L}( \Phi^\Sigma, \tilde D_\mu\Phi^\Sigma ,\ldots)\,,
\end{equation}
where $\Phi^\Sigma=l^\ast_\Sigma\Phi$,  $\tilde D=\dd +i\tilde \theta(\Sigma)$, and $\tilde\theta(\Sigma)=l^\ast_\Sigma\theta (\Sigma)$. Notice that the $1$-form  $\tilde \theta$ is nothing but the potential for the covariant strength tensor, $F^{\sfs}=\dd \tilde\theta$. Therefore, making the field redefinition $\Phi\rightarrow \Phi^\Sigma$, we obtain a local Lagrangian without higher derivatives. For instance,  a constant Poisson bivector $\pi^{\mu\nu}$ leads the  covariant derivative of the form
\begin{equation}
\begin{array}{c}
\tilde D_\mu \Phi^\Sigma=\partial_\mu \Phi^\Sigma+i(A_\mu+\frac12 \pi^{\lambda\nu}A_\lambda\partial_\mu A_\nu)\Phi^\Sigma\,,
\end{array}
\end{equation}
cf. (\ref{FF}). After integration over spacetime, the invariant and covariant Lagrangians give the same action functional. By adding the Lagrangians (\ref{Lcov}) and (\ref{tD}),  we arrive at a complete gauge theory which, being a deformation of ordinary electrodynamics, is capable of reproducing the effects of noncommutativity in a low-energy limit. 

To go beyond the low-energy approximation captured by Poisson electrodynamics, one needs a full noncommutative gauge theory with a prescribed semi-classical limit. At the kinematical level, this implies the construction of the corresponding noncommutative algebra by quantizing the underlying Poisson manifold. Again, the geometry of symplectic groupoids offers a natural way to do this. The idea is to obtain the quantum algebra as a twisted convolution $C^\ast$-algebra of a symplectic groupoid endowed with a suitable polarisation; see, e.g. \cite{Hawkins2008, Saemann:2012ab} and references therein. If successful, this will open up the prospect of constructing gauge theories on noncommutative spaces, more general than the Weyl--Moyal space.
We are going to come back to this issue elsewhere.

\section*{Acknowledgements} 

I am very grateful to Vladislav Kupriyanov and Richard Szabo for many stimulating conversations. The work was supported by the Ministry of Science and Higher Education of the Russian Federation (project No. FSWM-2020-0033).

\appendix

\section{Cohomology of groups} \label{App} To give the reader a broader algebraic perspective on the calculations in the main text, we provide some details about group cohomology.

Given a group $G$ that acts by linear transformations on a vector space $V$, consider functions $f_n(g_1,g_2,\ldots, g_n)$ on $n$ group elements with values in $V$. The functions form a linear space $C^n(G,V)$ and can be endowed with the action of a linear operator $\delta: C^n(G,V)\rightarrow C^{n+1}(G,V)$, which takes a function of $n$ variables $g$ to a function that depends on $n+1$ such variables: 
\begin{equation}
\begin{array}{ll}
  (\delta f_n)(g_1,\ldots, g_{n+1})&  \displaystyle = g_1f_n(g_2,\ldots,g_{n+1})+\sum_{k=1}^{n}(-1)^kf_n(g_1,\ldots, g_{k-1}, g_kg_{k+1},\ldots, g_{n+1})\\[5mm]
  &+(-1)^{n+1}f_n(g_1,\ldots, g_n)\,.
    \end{array}
\end{equation}
A simple calculation shows that $\delta^2=0$. This allows us to use the standard constructions and terminology of homological algebra. In particular, the function $f_n(g_1,\ldots,g_n)$ is called an $n$-cochain; a cochain $f_n=\delta f_{n-1}$ is a coboundary, whereas  a cochain $f_n$ satisfying $\delta f_n=0$ is called an $n$-cocycle. A cocycle is nontrivial if it is not a coboundary. The coboundaries constitute a subspace $B^n(G,V)\subset Z^n(G,V)$ in the space of all cocycles and the space of nontrivial cocycles is identified with the quotient  $H^n(G,V)=Z^n(G,V)/B^n(G,V)$
called the $n$-th cohomology group of $G$ with coefficients in $V$. In our interpretation of Eq. (\ref{sss}) as defining a $1$-cocycle of the group $\mathscr{L}(\mathcal{G})$ the role of the representation space $V$ is played by the space of maps  $f:\mathscr{B}(\mathcal{G})\rightarrow C^\infty(X)$ with the left action of 
$\mathscr{L}(\mathcal{G})$ defined by 
\begin{equation}
f(\Sigma)\rightarrow\Sigma_1 f(\Sigma):=f(\Sigma\Sigma_1)\,,\qquad \forall \Sigma_1\in \mathscr{L}(\mathcal{G})\,.
\end{equation}
The general $n$-cochain is represented by a function $f_n(\Sigma;\Sigma_1,\ldots,\Sigma_n)$ with $\Sigma\in\mathscr{B}(\mathcal{G})$ and the other arguments $\Sigma_i\in\mathscr{L}(\mathcal{G})$; for a given set of arguments, $f_n$ is a smooth function on $X$. We also consider more general possibilities, where the $f_n$'s assume values in the spaces of differential forms $\Lambda^k(X)$ of various degrees. To simplify our terminology and notation, we refer to such an $f_n$ as an $n$-cochain with values in $\Lambda^k(X)$ and denote the corresponding cochain complex and cohomology groups by ${C}^n(\mathcal{G},\mathsf{\Lambda}^k(X))$ and $H^n(\mathcal{G},\mathsf{\Lambda}^k(X))$, respectively.

\end{document}